# Algebra-Logical Repair Method for FPGA Logic Blocks

Vladimir Hahanov, Eugenia Litvinova, Wajeb Gharibi, Olesya Guz

*Abstract* – **An algebra-logical repair method for FPGA functional logic blocks on the basis of solving the coverage problem is proposed. It is focused on implementation into Infrastructure IP for system-on-a chip and system-in-package. A method is designed for providing the operability of FPGA blocks and digital system as a whole. It enables to obtain exact and optimal solution associated with the minimum number of spares needed to repair the FPGA logic components with multiple faults.**

*Index terms* – **digital system, system-on-a chip, system-in-package, FPGA functional logic blocks**.

## I. INTRODUCTION

At present there are many scientific publications, which cover SoC/SiP testing, diagnosis and repair problems [1-16, 19-20]. The testing and repair problem for the digital system logic components has a special place, because repair of faulty logic blocks is technologically complicated problem. Existing solutions, which are proposed in published works, can be divided on the following groups:

1. Duplication of logic elements or chip regions to double hardware realization of functionality. When faulty element is detected switching to faultless component by means of a multiplexer is carried out [4]. The FPGA models, proposed by Xilinx, can be applied for repair of Altera FPGA components. At repair the main unit of measure is row or column.

2. Application of genetic algorithms for diagnosis and repair on basis of off-line FPGA reconfiguration not using external control devices [5]. The fault diagnosis reliability is 99%, repair time is 36 msec instead of 660 sec, required for standard configuration of a project.

3. Time-critical FPGA repairing by means of replacement of local CLBs by redundant spares is proposed in [6,7]. In critically important applications the acceptable integration level for CLB replacement is about 1000 logic blocks.

The repair technologies for digital system logic, implemented on-chip FPGA, are based on existence or introduction of LUT redundancy after place and route procedure executing. Physical faults, which appear in the process of fabrication or operation, become apparent as logical or temporary failure and result in malfunction of a digital system. Faults are tied not only to the gates or LUT components but also to a specified location on a chip. The idea of digital system repairing comes to the removal of a fault element by means of repeated place and route executing after diagnosis. At that two repair technologies are possible: 1) Blockage of a defective area by means of development the control scripts for long time place and route procedure. But it is not always acceptable for real time digital systems. The approach is oriented on removal the defective areas of any multiplicity. Blockage of the defective areas by means of repeated place and route executing results in repair of a digital system. 2) Place and route executing for repairing of real time digital systems can result in disastrous effects. It is necessary the technological approach that allows repairing of the digital system functionality for milliseconds, required for reprogramming FPGA by new bitstream to remove defective areas from chip functionality. The approach is based on preliminary generation all possible bitstreams for blockage future defective areas by means of their logical relocation to the redundant nonfunctional chip area. The larger a spare area the less a number of bitstreams, which can be generated a priori. Concerning multiple faults, not covered by a spare area, it is necessary to segment a digital project by its decomposition on disjoin parts, which have their own Place and Route maps. In this case a digital system that has n spare segments for n distributed faults can be repaired. The total chip area consists of (n+m) equal parts.

The research objective is development of repair method for FPGA logic blocks on basis of use the redundant chip area.

Problems: 1) Development of an algebra-logical repair method for logic blocks of a digital system on basis of FPGA. 2) Development of a method for logic blocks matrix traversal to cover FPGA faulty components by spare tiles. 3) Analysis of practical results and future research.

Vladimir Hahanov is with the Kharkov National University of Radioelectronics, Ukraine, 61166, Kharkov, Lenin Prosp., 14, room 321 (corresponding author to provide phone: (057)7021326; fax: (057)7021326; e-mail: hahanov@ kture.kharkov.ua).
Eugenia Litvinova is with the Kharkov National University of Radioelectronics, Ukraine, 61166, Kharkov, Lenin Prosp., 14, room 378 (phone: (057)7021421; fax: (057)7021326; e-mail: kiu@ kture.kharkov.ua).
Wajeb Gharibi is with Kingdom of Saudi Arabia (e-mail: gharibiw2002@yahoo.com),
Olesya Guz is with the Donetsk Institute of Road Transport, Donetsk, Ukraine (e-mail: kiu@kture.kharkov.ua)

II. ALGEBRA-LOGICAL REPAIR METHOD FOR FPGA BLOCKS

The exact repair method for FPGA logic blocks by means of spares is represented. It enables to obtain quasi-optimal solution of coverage problem for faulty cells set by minimum quantity of spares. The method focuses on the implementation of digital system-in-package functionality to the Infrastructure IP. The objective function is minimization of FPGA spares $S^t$, needed for the repair of faulty blocks in the SiP operation by means of synthesis the disjunctive normal form of fault coverage and subsequent selection of a minimum conjunctive term $(S_1^t, S_2^t, ..., S_i^t, ..., S_{m_t}^t) \in S^t$ that answers the constraints on the number of spares $m_t \leq p$, which enter into the logical product:

$$Z = \min_{t=1,n}(|S^t|)|_{m_t \leq p}, S = (S^1 \vee S^2 \vee ... \vee S^t \vee ... \vee S^n); S^* = \{S_1, S_2, ..., S_j, ..., S_p\}.$$

FPGA functionality model is represented by a matrix of logic blocks operating by rows and columns of the structure $M = |M_{ij}|$. In the process of designing the matrix attached to a spare, consisting of rows and columns, which can be readdressed in the process of structure reconfiguration, when faults are detected.

A model of determining the minimum number of spares (rows and columns), covering all detected faults in a matrix of FPGA logic blocks, comes to the following items:

1. Making a coverage table for detected faulty FPGA blocks by spare rows and columns. To achieve the goal a topological model of testing results for FPGA functionality is considered in the form of matrix, coordinates of which identify detected faults (faultless and faulty blocks):

$$M = |M_{ij}|, M_{ij} = \begin{cases} 1 \leftarrow T \oplus f = 1; \\ 0 \leftarrow T \oplus f = 0. \end{cases} \quad (1)$$

Here the matrix coordinate is equal to 1 if modulo 2 sum of faultless behaviour for the block f and real test response give 1 value, which corresponds to the defect in a block. After FPGA testing and fixation all faulty blocks the construction of faults coverage table $Y = |Y_{ij}|, i = \overline{1,n}; j = \overline{1,m}$ is performed. Here columns correspond to a fault set ("1" coordinates), fixed in the matrix M ($|M| = m$), rows are numbers of columns and rows of FPGA blocks matrix, which cover faults, indicated in columns:

$$Y = |Y_{ij}|, Y_{ij} = \begin{cases} 1 \leftarrow C_i(R_i) \cap F_j \neq \varnothing; \\ 0 \leftarrow C_i(R_i) \cap F_j = \varnothing. \end{cases} \quad (2)$$

Operating by rows and columns, which contain faults, we can find an optimal solution in the form of fault coverage in the metrics of faulty rows and columns. Then the trivial reassignment procedure is carried out for faulty rows and columns, which replaced by faultless FPGA spare components.

To improve the solving of coverage task one-dimensional vector, concatenated from two sequences C and R by power $n = p + q$, is used instead of two-dimensional metrics components:

$$X = C * R = (C_1, C_2, ... C_i, ... C_p) * (R_1, R_2, ... R_j, ... R_q) = \\ = X^c * X^r = (X_1, X_2, ... X_i, ... X_p, X_{p+1}, X_{p+2}, ... X_{p+j}, ... X_{p+q}). \quad (3)$$

At that there is one-to-one correspondence between elements of initial sets (C, R) and resultant vector X that is established in the first column of Y matrix. Transformation $X = C * R$ is carried out for ease of consideration and subsequent construction the disjunctive normal form in the frame of uniformity in the variables, which form the Boolean function. If given procedure is not performed, a function will be defined by variables of two types, which consist rows and columns of a memory matrix.

2. CNF forming by analytic, complete and exact solving of the coverage task. After the forming of a coverage matrix that contains zero and unit coordinates, the synthesis of coverage analytic form by writing CNF for unit coordinates of matrix columns is carried out. A number of conjunctive terms is equal to quantity of table columns, and the length of a disjunctive term is equal to quantity of "1" in the column:

$$Y = \bigwedge_{j=1}^{m} (Y_{pj} \vee Y_{qj})\{Y_{pj}, Y_{qj}\} = 1 = \bigwedge_{j=1}^{m} (X_{pj} \vee X_{qj}). \quad (4)$$

Last expression shows that every column identifies two variants of fault coverage – rows and columns. So a column has only two coordinates, which have unit value, and numbers of logical products is equal to total quantity of faults m, detected in FPGA matrix.

3. Transformation CNF to DNF enables to obtain all solutions of the coverage task. For this it is necessary to apply the logical product operation and minimization rules to the conjunctive normal form to get disjunctive normal form:

$$Y = \bigvee_{j=1}^{w} (k_1^j X_1 \wedge k_2^j X_2 \wedge ... \wedge k_i^j X_i \wedge ... \wedge k_n^j X_n), k_i^j = \{0,1\}. \quad (5)$$

The generalize DNF is represented here; a number of terms is equal to $w = 2^n$, n – numbers of rows in the set (C,R) or quantity of variables X in the matrix Y. All possible solutions (fault coverages by spares) are written by the set of row identifiers for a coverage table. If $k_i^j$ at $X_i$ is equal to zero, $X_i$ is changed to an insignificant variable.

4. Choice the minimum and exact solutions of the coverage task. The procedure involves the determination of minimum length conjunctive terms by Quine in the DNF. Its subsequent transformation to the rows and columns of a memory matrix on basis of earlier correspondence enables to write the minimum coverage (or a set of them) by two-dimensional row and column metrics that satisfies the constraints of the objective function on spare quantity.

5. Realization of reassignment procedure for faulty rows and columns by similar faultless components of FPGA spare.

Example 2.2. Fulfill the repair process for FPGA matrix to determine the minimum quantity of spares, covering all faults. The matrix with faults and spares, highlighted by black color [19,20], is shown in Fig. 1.

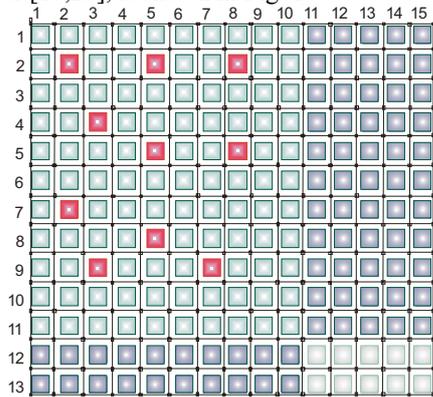

Fig. 1. FPGA with faults and spares

The matrix has a spare for diagnosis and repair of faulty cells that is two rows and five columns. According to item 1 of the model for determining the minimum quantity of spares, covering all detected faults of a memory matrix, a coverage table for 10 faults

$$F = (F_{2,2}, F_{2,5}, F_{2,8}, F_{4,3}, F_{5,5}, F_{5,8}, F_{7,2}, F_{8,5}, F_{9,3}, F_{9,7})$$

and eleven rows is formed:

| $X_i / F_{i,j}$ | $F_{2,2}$ | $F_{2,5}$ | $F_{2,8}$ | $F_{4,3}$ | $F_{5,5}$ | $F_{5,8}$ | $F_{7,2}$ | $F_{8,5}$ | $F_{9,3}$ | $F_{9,7}$ |
|---|---|---|---|---|---|---|---|---|---|---|
| $C_2 \to X_1$ | 1 | | | | | | 1 | | | |
| $C_3 \to X_2$ | | | | 1 | | | | | 1 | |
| $C_5 \to X_3$ | | 1 | | | 1 | | | 1 | | |
| $C_7 \to X_4$ | | | | | | | | | | 1 |
| $C_8 \to X_5$ | | | 1 | | | 1 | | | | |
| $R_2 \to X_6$ | 1 | 1 | 1 | | | | | | | |
| $R_4 \to X_7$ | | | | 1 | | | | | | |
| $R_5 \to X_8$ | | | | | 1 | 1 | | | | |
| $R_7 \to X_9$ | | | | | | | 1 | | | |
| $R_8 \to X_{10}$ | | | | | | | | 1 | | |
| $R_9 \to X_{11}$ | | | | | | | | | 1 | 1 |

(6)

Power or the number of rows in the table is determined by concatenation of columns C and rows R, which are in the one-to-one correspondence with the vector of variables X:

$$C*R = (C_2, C_3, C_5, C_7, C_8)*(R_2, R_4, R_5, R_7, R_8, R_9) \approx$$
$$\approx X = (X_1, X_2, X_3, X_4, X_5, X_6, X_7, X_8, X_9, X_{10}, X_{11}). \quad (7)$$

Construction of CNF is performed by a coverage table by means of writing the terms for unit values of columns:

$$Y = (X_1 \vee X_6)(X_3 \vee X_6)(X_5 \vee X_6)(X_2 \vee X_7)(X_3 \vee X_8) \& \\ \& (X_5 \vee X_8)(X_1 \vee X_9)(X_3 \vee X_{10})(X_2 \vee X_{11})(X_4 \vee X_{11}). \quad (8)$$

Getting of disjunctive normal form is based on the Boolean algebra identities, which allows performing the logical multiplication all ten multiplicands and the subsequent minimization of DNF terms by applying the operator $ab \vee a\bar{b} = a$, absorption axioms and removing the same terms.

The final result is:

$$Y = X_1X_2X_3X_4X_5 \vee X_2X_3X_4X_5X_6X_9 \vee X_1X_2X_3X_4X_6X_8 \vee \\ \vee X_2X_3X_4X_6X_8X_9 \vee X_1X_2X_4X_6X_8X_{10} \vee X_2X_4X_6X_8X_9X_{10} \vee \\ \vee X_1X_4X_6X_7X_8X_{10}X_{11} \vee X_1X_2X_3X_5X_{11} \vee X_2X_3X_5X_6X_9X_{11} \vee \\ \vee X_1X_2X_3X_6X_8X_{11} \vee X_2X_3X_6X_8X_9X_{11} \vee X_1X_2X_6X_8X_{10}X_{11} \vee \\ \vee X_2X_6X_8X_9X_{10}X_{11} \vee X_1X_3X_5X_7X_{11} \vee X_3X_5X_6X_7X_9X_{11} \vee \\ \vee X_1X_3X_6X_7X_8X_{11} \vee X_3X_6X_7X_8X_9X_{11} \vee X_1X_6X_7X_8X_{10}X_{11} \vee \\ \vee X_6X_7X_8X_9X_{10}X_{11}. \quad (9)$$

The choice of minimum length terms with 5 variables in given case determines a set of minimal solutions:

$$Y = X_1X_2X_3X_4X_5 \vee X_1X_2X_3X_5X_{11} \vee X_1X_3X_5X_7X_{11} \quad (10)$$

Transformation of the function to the coverage that contains variables in the form of FPGA rows and columns enables to represent the terms as follows:

$$Y = C_2C_3C_5C_7C_8 \vee C_2C_3C_5C_8R_9 \vee C_2C_5C_8R_4R_9. \quad (11)$$

All minimum solutions satisfy the requirements on quantity of spares, determined by numbers:

$$(|C^r| \leq 5) \& (|R^r| \leq 2).$$

The subsequent repair technology for faulty FPGA blocks is electrical reprogramming of an address decoder for FPGA column or row. Concerning FPGA shown in Fig. 1 the readdressing of columns with faulty logic blocks to spare columns is realized, for instance, in compliance with first term of (11) that defines the relation:

| Faulty column | 2 | 3 | 5 | 7 | 8 |
|---|---|---|---|---|---|
| Spare column | 11 | 12 | 13 | 14 | 15 |

The computational complexity of an algebra-logical repair method for solving of the coverage task [17,20] is determined by the following expression:

$$Q = 2^{|F|} + |C+R| \times 2^{|F|}, \quad (12)$$

where $2^{|F|}$ – costs related to the synthesis of DNF by means of logical multiplication two-component disjunctions, the number of which is equal to quantity of faulty blocks (the fault coordinate is determined by a row or column number); $|C+R| \times 2^{|F|}$ is upper limit of computational cost needed to minimize the DNF on limiting set of variables that is equal to total quantity of rows and columns $|C+R|$.

In the worst case, when the coordinates of all faulty blocks are not correlated by rows and columns (unique), for instance, diagonal faults, computational complexity of the matrix method becomes dependent on the number of faulty cells only, and its analytic form is transformed to the following expression:

$$Q = 2^{|F|} + |C+R| \times 2^{|F|}\big|_{|C+R| \leq 2 \times |F|} = \\ = 2^{|F|} + 2 \times |F| \times 2^{|F|} = 2^{|F|} \times (1 + 2 \times |F|). \quad (13)$$

If to use the numbers of faults m instead of the power of a fault set the previous expression is represented in simpler form:

$$Q = 2^m \times (1 + 2 \times m) = 2^m(2m+1). \quad (14)$$

According to the Functional Intellectual Property, an algebra-logical repair method for FPGA logic blocks on basis of solving the coverage task is implemented into a chip as one of Infrastructure IP components that is designed for support of FPGA blocks and SoC availability.

### III. GALLS METHOD FOR A LOGIC BLOCK MATRIX TO COVER THE FAULTY FPGA COMPONENTS BY SPARE TILES

Topology of a chip is represented by the tile matrix $M = |M_{ij}| i = \overline{1,p}; j = \overline{1,q}$ scalable horizontal and vertical by integers ($p \times q$). Every tile $M_{ij}$ has $n^2$ logic blocks. The matrix has an arbitrary number of faults, equal to k. There are less or equal $n^2$ faulty logic blocks in every tile. An example of a tile matrix with faults is shown in Fig. 2. Here the dimension of tile n is equal to 3 and matrix dimension in the number of tiles by rows and columns is equal to 5.

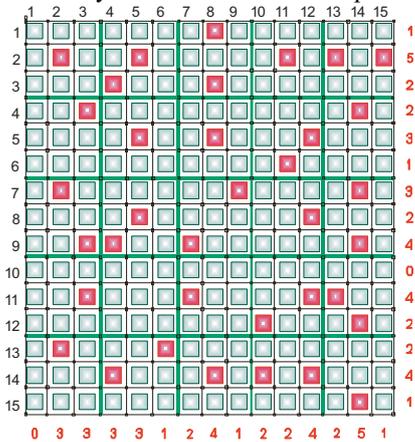

Fig. 2. FPGA block matrix in the number of tiles

The traversal method for a matrix of logic blocks is represented by the steps below. It is designed to repair the FPGA components and enables to get quasi-optimal coverage of all faulty blocks by minimum numbers of spare tiles.

1. Determination of coordinates for all faulty blocks of the matrix $M = |M_{ij}|$, specifies the topology of a chip.

2. Construction of binary coverage matrixes for faulty blocks by traversal of the tiles in rows and columns, which dimension is determined by the parameters respectively:

$$M_r = |M_{ij}^r|, i = \overline{1,p/n}; j = \overline{1,q};$$

$$M_c = |M_{ij}^c|, i = \overline{1,p}; j = \overline{1,q/n}.$$

Here every n coordinates of a row (column) is replaced by one coordinate with value that is determined by $f^r(f^c)$ function Or of n coordinates.

$$M_{ij}^c = f^r(f^c) = \begin{cases} 0 \leftarrow (000); \\ 1 \leftarrow (1XX) \vee (X1X) \vee (XX1), X = \{0,1\}. \end{cases}$$

For instance, a procedure for obtaining a row traversal matrix gives the result

$$M = |M_{ij}| = \begin{vmatrix} 0 & 0 & 0 & 0 & 1 & 1 \\ 0 & 0 & 0 & 0 & 0 & 1 \\ 0 & 1 & 1 & 0 & 1 & 0 \\ 1 & 0 & 0 & 0 & 0 & 1 \\ 0 & 0 & 0 & 0 & 1 & 0 \\ 0 & 1 & 0 & 1 & 0 & 1 \end{vmatrix} \xrightarrow{f^r} M_r = |M_{ij}^r| = \begin{vmatrix} 0 & 1 & 1 & 0 & 1 & 1 \\ 1 & 1 & 0 & 1 & 1 & 1 \end{vmatrix}.$$

Here each column is compressed in the two coordinates on the rules of logical operation Or, because the tile parameter n is equal to 3 here and below.

Similarly a procedure for obtaining a column traversal matrix gives the result

$$M = |M_{ij}| = \begin{vmatrix} 0 & 0 & 0 & 0 & 1 & 1 \\ 0 & 0 & 0 & 0 & 0 & 1 \\ 0 & 1 & 1 & 0 & 1 & 0 \\ 1 & 0 & 0 & 0 & 0 & 1 \\ 0 & 0 & 0 & 0 & 1 & 0 \\ 0 & 1 & 0 & 1 & 0 & 1 \end{vmatrix} \xrightarrow{f^c} M_c = |M_{ij}^c| = \begin{vmatrix} 0 & 1 \\ 0 & 1 \\ 1 & 1 \\ 1 & 1 \\ 0 & 1 \\ 1 & 1 \end{vmatrix}.$$

3. Determination of coverage quality criteria for faulty blocks by using the binary matrixes on basis of counting the numbers of unit coordinates reduced to the actual unit matrix.

The row coverage criterion faulty blocks is determined by the following expression:

$$Q_r = \sum_{i=1}^{p/n} \left[ \frac{1}{H_i^r - L_i^r + 1} \sum_{j=1}^{q} M_{ij}^r \right].$$

Here $H_i^r(L_i^r)$ – maximum (minimum) index of j-th coordinate for a row of the matrix $|M_{ij}^r|$, after (before) which there are only zero coordinates in a row. Actually $H_i^r - L_i^r + 1$ is the units spread interval in rows of the matrix $|M_{ij}^r|$, which gives the sum of unit row coordinates. Further reduced estimates for all rows are added, which is the criterion of row coverage for faulty blocks.

The column coverage criterion is the following:

$$Q_c = \sum_{j=1}^{q/n} \left[ \frac{1}{H_j^c - L_j^c + 1} \sum_{i=1}^{p} M_{ij}^c \right].$$

Here $H_j^c(L_j^c)$ – maximum (minimum) index of j-th coordinate for a column of the matrix $|M_{ij}^c|$, after (before) which there are only zero coordinates in a column. Actually $H_j^c - L_j^c + 1$ is the units spread interval in columns of the matrix $|M_{ij}^c|$, which gives the sum of unit column coordinates. Further reduced estimates for all columns are added, which is the criterion of column coverage for faulty blocks.

4. The decision on the choice of strategy $S = \{S_r, S_c\}$ for coverage of faulty logic blocks by spares by means of comparing the values of the structurization criteria $Q_r, Q_c$ for the rows and columns:

$$S = \begin{cases} S_r \leftarrow Q_r < Q_c; \\ S_c \leftarrow Q_r \geq Q_c. \end{cases}$$

In the first case, the faulty blocks coverage strategy is realized by sequential traversal of all tile rows. The second one – the traversal of tile columns is done.

5. The tile traversal strategy by rows is realized by the modified matrix $\left|M_{ij}^r\right|$. Each matrix row is represented by the binary vector $M_i^r = (M_{i1}^r, M_{i1}^r, ..., M_{ij}^r, ..., M_{iq}^r)$. Step 1. Nulling of a zero coordinate counter and a spare tile counter: $j = 0$, $Q = 0$. Step 2. Sequential scanning of the vector elements $j = j+1 \leftarrow M_{ij}^r = 0$ up to the first "1" encountered $M_{ij}^r = 1 \rightarrow (Q = Q+1, j = j+n-1)$. From this "1" it is counted n tiles, which are covered by the spare tile. The number of spare tiles Q is increased by 1. Step 3. If the condition $j \geq q$ is true – the end of row processing. Otherwise – go to step 2. The procedure is applied to all rows of the modified matrix $\left|M_{ij}^r\right|$ ($\left|M_{ij}^c\right|$), resulting in counter Q will contain the minimum number of spares to cover all faulty blocks. Similarly the tile traversal strategy by columns ($\left|M_{ij}^c\right|$) is performed. In this case the index i is changed instead of j in a traversal procedure.

6. Definition of the coverage quality for obtained solution by means of counting the number of faulty blocks of the matrix, reduced to the minimum number of spares N, covering all faulty blocks:

$$Q_{cr} = \frac{1}{N}\sum_{i=1}^{n} F_i.$$

7. The end of finding of a quasioptimal cover of faulty blocks by spares.

Example. For FPGA, shown in Fig. 2, according to item 2 of the repair model the construction of two matrixes is performed:

$$M_r = \left|M_{ij}^r\right| = \begin{array}{|c|c|c|c|c|c|c|c|c|c|c|c|c|c|c|} \hline 0 & 1 & 0 & 1 & 1 & 0 & 0 & 1 & 0 & 0 & 1 & 0 & 1 & 0 & 1 \\ \hline 0 & 0 & 1 & 0 & 1 & 0 & 0 & 1 & 0 & 0 & 1 & 1 & 0 & 1 & 0 \\ \hline 0 & 1 & 1 & 1 & 1 & 0 & 1 & 0 & 1 & 0 & 0 & 1 & 0 & 1 & 0 \\ \hline 0 & 0 & 1 & 0 & 0 & 0 & 1 & 0 & 0 & 1 & 0 & 1 & 1 & 1 & 0 \\ \hline 0 & 1 & 0 & 1 & 0 & 1 & 0 & 1 & 0 & 1 & 0 & 1 & 0 & 1 & 0 \\ \hline \end{array}$$

$$M_c = \left|M_{ij}^c\right| = \begin{array}{|c|c|c|c|c|} \hline 0 & 0 & 1 & 0 & 0 \\ \hline 1 & 1 & 0 & 1 & 1 \\ \hline 0 & 1 & 1 & 0 & 0 \\ \hline 1 & 0 & 0 & 0 & 1 \\ \hline 0 & 1 & 1 & 1 & 0 \\ \hline 0 & 0 & 0 & 1 & 0 \\ \hline 1 & 0 & 1 & 0 & 1 \\ \hline 0 & 1 & 0 & 1 & 0 \\ \hline 1 & 1 & 1 & 0 & 1 \\ \hline 0 & 0 & 0 & 0 & 0 \\ \hline 1 & 0 & 1 & 1 & 1 \\ \hline 0 & 0 & 0 & 1 & 1 \\ \hline 1 & 1 & 0 & 0 & 0 \\ \hline 0 & 1 & 1 & 1 & 0 \\ \hline 0 & 0 & 0 & 0 & 1 \\ \hline \end{array}$$

Further, in accordance with paragraph 3, the calculation of the structurization criteria is performed for the matrix above:

$$Q_r = \sum_{i=1}^{p/n}\left[\frac{1}{H_i^r - L_i^r + 1}\sum_{j=1}^{q}M_{ij}^r\right] = \frac{7}{14} + \frac{6}{12} + \frac{8}{13} + \frac{6}{12} + \frac{7}{13} = 2{,}64;$$

$$Q_c = \sum_{j=1}^{q/n}\left[\frac{1}{H_j^c - L_j^c + 1}\sum_{i=1}^{p}M_{ij}^c\right] = \frac{6}{12} + \frac{7}{13} + \frac{7}{14} + \frac{7}{13} + \frac{7}{14} = 2{,}58.$$

As it can be seen from the above-mentioned criteria, the number of faulty coordinates in the columns reduced to the unit matrix is less than in the rows. Therefore, subject to item 4, the strategy for solving the coverage problem by column traversal is chosen: $S_c \leftarrow (Q_r = 2{,}64 \geq Q_c = 2{,}58)$.

Structural appeal of columns higher then rows, because the number of faults reduced to the matrix less than in the first case.

This strategy gives the actual quality of coverage – the number of faulty logic blocks of the matrix (by one spare tile), reduced to the necessary quantity of spares that is equal to:

$$Q_c^* = \frac{1}{N}\sum_{i=1}^{n} F_i = 36/20 = 1{,}8.$$

For comparison – the row traversal procedure gives a lower quality:

$$Q_r^* = \frac{1}{N}\sum_{i=1}^{n} F_i = 36/21 = 1{,}71.$$

The final coverage of a fault set has one tile more (21) than the solution obtained by the first method (20). So, the choice of coverage strategy on the basis of calculation and comparison the criteria for counting the number of unit coordinates, reduced to the actual unit matrix, confirms their consistency and the subsequent optimality of obtained coverage.

Fig. 3 shows statistics of the processing of different types logic block matrixes with faulty components. It demonstrates correctness of applying the proposed criterion, which shows the optimal and efficient strategy of matrix traversal to obtain minimum coverage of faulty logic blocks by spare tiles in all cases except the last one. In the latest variant the structurization criteria by rows greater than by columns $Q_r = 1{,}78 > Q_c = 1{,}55$ that is the basis of covering the faults by columns. In this case the coverage quality is $Q_c^* = 1{,}125$ that corresponds to 16 spare tiles needed to repair all 18 faulty blocks. While the optimal solution is traversal of the matrix by columns, where the coverage quality is $Q_c^* = 1{,}2$ that corresponds to 15 spare tiles only.

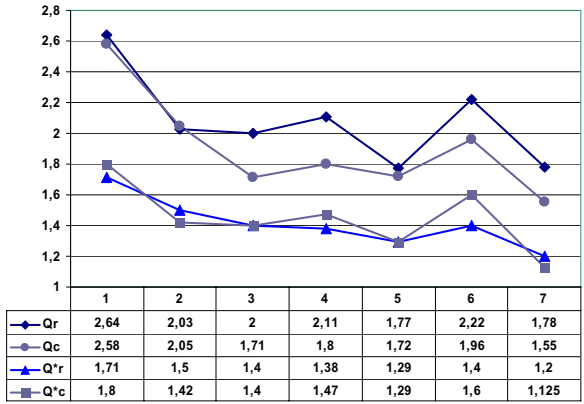

Fig. 3. Structurization criteria and coverage quality for the examples

IV. SOFTWARE "aGALLS" FOR COVERING OF FAULTY FPGA COMPONENTS BY SPARE TILES

Software «aGalls» is designed for covering of faulty FPGA components by spare tiles by using the traversal method for logic block matrix.

The initial information is represented by a matrix of tiles $M = |M_{ij}| i = \overline{1,p}; j = \overline{1,q}$, the dimension $p \times q$, where each tile has n*n logic blocks. The initial quantity of faults k is made to the matrix. The software detects the coordinates of all faults, constructs the binary coverage matrixes for faulty blocks by means of traversal the tiles by columns and rows, and calculates the quality criteria of fault coverage (Fig. 4).

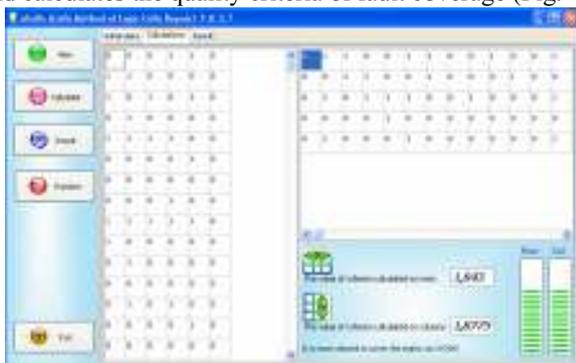

Fig. 4. Calculation results

The program provides the opportunity to review the changes of temporary matrixes after covering the faulty blocks both horizontally and vertically. In addition, the actual number of spare tiles needed to cover all faults is displayed; it enables to estimate the correctness of the selected solution on the basis of calculating the criteria (Fig. 5).

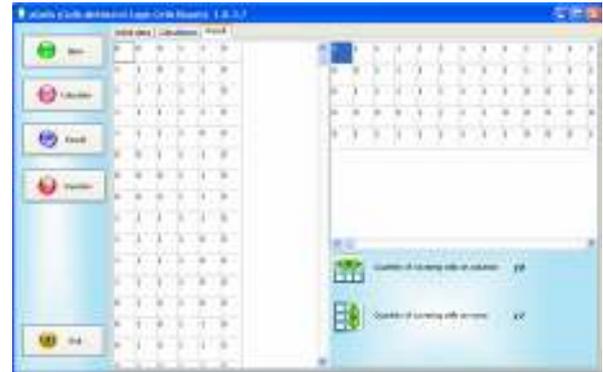

Fig. 5. Coverage of a matrix with faults and spares

An algorithm for solving the coverage problem implemented in the software is represented below and in Fig. 6.

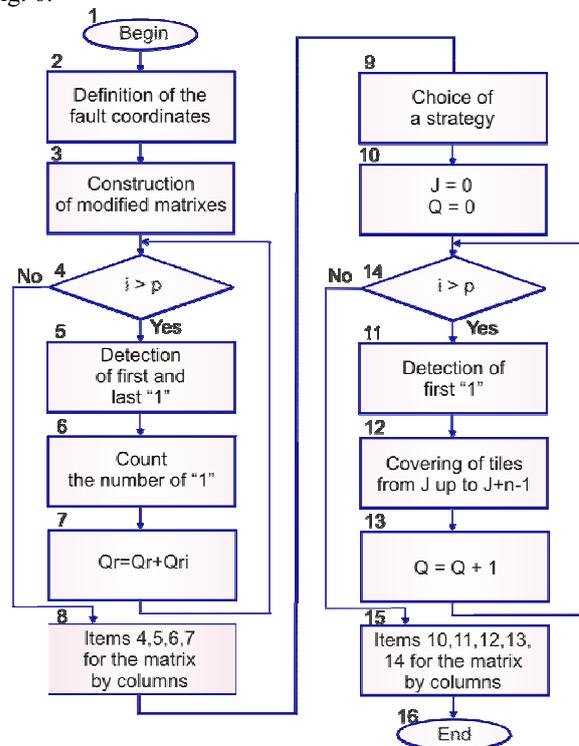

Fig. 6. Model of solving the optimal coverage task

1. Beginning of solving the coverage problem for a logic block matrix by spare tiles.

2. Determination of the coordinates for all faulty blocks of the matrix.

3. Construction of the binary coverage matrixes for faulty blocks to traverse the tiles by rows and columns.

4. Construction the fault coverage for the modified matrix by means of their traversal by rows.

5. Detection the first and last "1".

6. Counting the number of "1" in a row.

7. Increment number of rows processed.

8. Calculation of the structurization criteria by rows.

9. Implementation of items 4, 5, 6, 7 for columns.

10. The decision on the choice of faulty blocks coverage strategy in the direction corresponding to the lower value of the criterion.

11. Nulling of zero coordinates counter and spare tile counter $j = 0, Q = 0$.

12. Sequential scanning of the vector elements $j = j+1 \leftarrow M_{ij}^r = 0$ up to the first "1" $M_{ij}^r = 1 \rightarrow (Q = Q+1, j = j+n-1)$ by rows.

13. From this "1" it is counted n tiles, which are covered by the spare.

14. The value Q is incremented by 1.

15. If $j \geq q$ – the end of row proceeding, otherwise – item 11.

16. Implementation of items 10, 11, 12, 13, 14 at traversal by columns. The end of searching the quasioptimal coverage of faulty blocks by spare tiles.

To check the validity of the criteria 200 experiments for the matrix M was carried out; the dimension of matrix is p*q, the number of blocks in the tile is n horizontally and vertically, and the number of faults is k, where

$p = \overline{3,7}; n = \overline{2,5}; q = \overline{3,7}; k = \overline{3,(n \times p \times q)}$.

As a result, by using the criteria the most efficient way to repair has not been determined in 29% of cases, respectively a positive result – 71%. Fig. 7 presents diagrams showing the values of structurization criteria and coverage quality at the traversal of the modified matrix by rows and columns for the 10 examples. The dimension of the matrix was constant: 4 tiles in the horizontal and 5 ones vertical, the dimension of the tiles was 3*3 blocks.

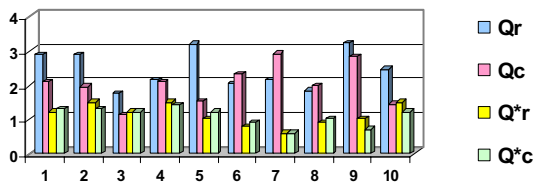

Fig. 7. Structurization criteria and coverage quality

The software is used to verify and test the method for obtaining the quasioptimal fault coverage by minimum quantity of spare tiles. It can be implemented into a chip as embedded Infrastructure IP component, where the initial information about the topology of a chip has to be memorized during Place and Route process.

V. CONCLUSION

Algebra-logical repair method for FPGA logic blocks on the basis of solving the coverage task is focused on the implementation into a chip as one of Infrastructure IP components. It is designed to repair the FPGA blocks and SiP as a whole.

The method enables to obtain exact and optimal solution associated with the minimum number of spare blocks needed to repair the FPGA logic components with multiple faults.

The traversal method for a logic block matrix is designed to repair the FPGA components by obtainment the solution in the form of quasioptimal coverage for all faulty blocks by minimum number of spare tiles. A choice one of two traversal strategies for rows or columns of a logic block matrix on the basis of the structurization criteria, which determine a number of faulty blocks, reduced to the unit modified matrix of rows or columns.

The technological solutions represented in the survey and proposed methods for diagnosis and repairing of digital system-on-a-chip and system-in-package correlates well with the analytical market research of electronics in 2009 and is formulated in the form of Gartner's Top 10 Strategic Technologies for 2009: 1) Virtualization. 2) Cloud Computing. 3) Servers – Beyond Blades. 4) Web-Oriented Architectures. 5) Enterprise Mashups. 6) Specialized Systems. 7) Social Software and Social Networking. 8) Unified Communications. 9) Business Intelligence. 10) Green IT [http://www.gartner.com/].

As well as Gartner's Top 10 correlates well with the analytical research of Computer Sciences Corporation (CSC), represented as 7 tendencies: 1) New media. Internet has become a full-fledged framework for creating and using audio, video and text content in a planetary scale. 2) Social software. Social networks attract millions of users, using the common interests. 3) Enhanced Reality. Gradually, but persistently it enters our lives. Virtual reality, where images of the users travel through the virtual worlds, it becomes practical in finding suitable products, services, products without their prior purchase. 4) Transparency of information. It will let you see yourself and the world with a given degree of detail by means of sensors and internet-cameras placed in the office, and throughout the world. The other side of the coin is how to hide and preserve personal space. 5) Wireless innovations. They allow running any application on any device, anywhere in the world. Here we should expect the appearance of conflict related to the division of frequencies between telecommunications operators, radio and television, cable and satellite companies, Internet service providers. It is expected the integrated solution of the problem on the basis of wireless technologies with mobile Internet-services. 6) New hardware and software platforms. Level of virtualization is increased. The number of applications which function on the same computer under different operating systems is growing exponentially. «Cloud computations», when a consumer pays for the use of computer infrastructure and applications to providers, stored client data on your own servers, significantly alter the entire calculation structure. Prospects of nanotechnologyб molecular, quantum and optical computing become more realistic. Instead of silicon chips lighter and smaller elements: the atoms, DNA, spins of electrons and light will work. 7) Intelligent world. Semantic and networking technologies allow computing devices to interpret the information on the algorithms of natural intelligence, whether text, voice, image, or life situations. Computers will be to teach, give advices, make predictions based on information received from the environment and the individual person. Self-learning semantic retrieval applications is developed in the Internet. A truly intelligent world is made, where people and computers will be able to talk and communicate with each other based on a combination of semantic and network technologies. This will result in the appearance of artificial intelligence, and possibly hyperintelligence, reading human thoughts [http://www.pcweek.ru/].